\documentclass[letter]{aa}
\usepackage{graphicx}
\usepackage{txfonts}
\begin{document}
\title{$J$, $H$, $K$ spectro-interferometry of the Mira variable S~Orionis
\thanks{Based on observations made with the VLT Interferometer (VLTI)
at Paranal Observatory under program ID 080.D-0691}}
\author{
M.~Wittkowski\inst{1} \and
D.~A.~Boboltz\inst{2} \and
T.~Driebe\inst{3} \and
J.-B.~Le~Bouquin\inst{4}
F.~Millour\inst{3}
K.~Ohnaka\inst{3} \and
M.~Scholz\inst{5,6}
}
\institute{
ESO, Karl-Schwarzschild-Str. 2,
85748 Garching bei M\"unchen, Germany,
\email{mwittkow@eso.org}
\and
US Naval Observatory, 3450 Massachusetts Avenue,
NW, Washington, DC 20392-5420, USA
\and
Max-Planck-Institut f\"ur Radioastronomie,
Auf dem H\"ugel 69, 53121 Bonn, Germany
\and
ESO, Casilla 19001, Santiago 19, Chile
\and
Institut f\"ur Theoretische Astrophysik der Univ. Heidelberg,
Albert-Ueberle-Str. 2, 69120 Heidelberg, Germany \and
Institute of Astronomy, School of Physics,
University of Sydney, Sydney NSW 2006, Australia
}
\date{Received \dots; accepted \dots}
\abstract{}
{We present $J$, $H$, $K$ spectrally dispersed interferometry with 
a spectral resolution of 35 for the Mira variable S~Orionis. We aim 
at measuring the diameter variation as a function of wavelength
that is expected due to molecular layers lying above the continuum-forming 
photosphere. Our final goal is a better understanding of the pulsating atmosphere
and its role in the mass-loss process.}
{Visibility data of S~Ori were obtained at phase 0.78 with the VLTI/AMBER
instrument using the fringe tracker FINITO 
at 29 spectral channels between 1.29\,$\mu$m and 2.32\,$\mu$m.  
Apparent uniform disk (UD) diameters were computed for each
spectral channel.
In addition, the visibility data were directly compared to predictions by recent
self-excited dynamic model atmospheres.}
{S~Ori shows significant variations in the visibility values
as a function of spectral channel that can only be described by a
clear variation in the apparent angular size with wavelength.
The closure phase values are close to
zero at all spectral channels, indicating the absence of asymmetric
intensity features.
The apparent UD angular diameter 
is smallest at about 1.3\,$\mu$m and 1.7\,$\mu$m and increases by
a factor of $\sim$\,1.4 around 2.0\,$\mu$m.
The minimum UD angular diameter at near-continuum wavelengths
is $\Theta_\mathrm{UD}$=8.1$\pm$0.5\,mas, 
corresponding to $R\sim$\,420\,R$_\odot$.
The S~Ori visibility data and the apparent UD variations 
can be explained reasonably well by a dynamic atmosphere model 
that includes molecular layers, particularly water vapor and CO. 
The best-fitting photospheric angular diameter 
of the model atmosphere is $\Theta_\mathrm{Phot}$=8.3$\pm$0.2\,mas,
consistent with the UD diameter measured at near-continuum wavelengths.}
{The measured visibility and UD diameter variations with wavelength
resemble and generally confirm the predictions by recent dynamic 
model atmospheres. 
These size variations with wavelength can be understood as the effects 
from water vapor and CO layers lying above the continuum-forming photosphere.
The major remaining differences between observations
and model prediction are very likely due to an 
imperfect match of the phase and cycle combination
between observation and available models.}
\maketitle
\section{Introduction}
Mira stars are low-mass, large-amplitude,
long-period variable stars on the AGB, evolving toward the planetary nebula 
and white dwarf phases. They exhibit a mass-loss rate
on the order of $\sim$10$^{-6}$\,M$_\odot$/year that significantly affects the 
further stellar evolution and is one of the most important
sources for the chemical enrichment of the interstellar medium. The
dust condensation sequence, the wind-driving mechanism, and the role of 
pulsation are currently not well understood, in particular for oxygen-rich AGB 
stars (Woitke et al. \cite{woitke06}; H\"ofner \& Andersen \cite{hoefner07}).
The pulsating atmospheres of Mira stars can become very extended
because of dynamic effects including shock fronts, and they are very cool in 
their outer parts. Here, 
molecules can form, which for O-rich stars are most 
importantly H$_2$O, CO, TiO, and SiO
(Tsuji et al. \cite{tsuji97}; Tej et al. \cite{tej03}; Ohnaka \cite{ohnaka04}).
Wittkowski et al. (\cite{wittkowski07}) found for the case of S~Ori
that Al$_2$O$_3$ dust condenses within the extended atmosphere at 
phase-dependent distances of 1.8--2.4 photospheric radii. This extended 
atmosphere, which is characterized by phase-dependent temperature
and density stratifications, the presence
of molecular layers, and the formation of dust, is thus of particular interest
for our better understanding of pulsation and mass loss.
Observed radii of Mira stars have been found to differ for different 
optical and infrared bandpasses (e.g. Thompson et al. \cite{thompson02};
Mennesson et al. \cite{mennesson02}; Ireland et al. \cite{ireland04a};
Perrin et al. \cite{perrin04}; Eisner et al. \cite{eisner07}), and this has been 
attributed to the presence of molecular layers located above
the continuum-forming photosphere.
Here, we present both a spectro-interferometric observation 
of the Mira star S~Ori that covers the near-infrared $J$, $H$, and $K$ bands 
simultaneously at a spectral resolution of 35 and a comparison to
recent self-excited dynamic model atmospheres.

S~Ori is a Mira variable star with spectral type M6.5e--M9.5e
and $V$ magnitude 7.2--14.0 (Samus et al. \cite{samus04}).
We use a Julian Date of last maximum brightness $T_0=2453190\ \mathrm{days}$, 
a period $P=430\ \mathrm{days}$ (as in Wittkowski et al. \cite{wittkowski07}), 
and the distance of
$480\ \mathrm{pc}\ \pm\ 120\ \mathrm{pc}$ from 
van Belle et al. (\cite{vanbelle02}).
The broadband near-infrared $K$ UD angular diameter
of S~Ori has been measured by van Belle et al. (\cite{vanbelle96}),
Millan-Gabet et al. (\cite{millan05}), and
Boboltz \& Wittkowski (\cite{boboltz05})
to values between 9.6\,mas and 10.5\,mas at different
phases. Joint VLTI/MIDI and VLBA/SiO maser observations by 
Wittkowski et al. (\cite{wittkowski07}) have shown that S~Ori
exhibits significant phase-dependencies of the atmospheric extension and
dust shell parameters with photospheric angular diameters between 7.9\,mas
and 9.7\,mas. 
\begin{table*}
\caption{Observation log. Night starting 12 October 2007, JD 2454386.}
\label{tab:obslog}
\begin{tabular}{lllllllrrrr}
\hline\hline
Target & Purpose & $\Theta_\mathrm{LD}$& DIT & Time & $\Phi_\mathrm{Vis}$ & $B_p$
[m]
& PA$_p$ & AM & Seeing & $\tau_0$\\
  &  & [mas] &[msec] &[UTC] &     & E0-G0/G0-H0/E0-H0   &$\deg$  && [$\arcsec$]&[msec]\\\hline
\object{45~Eri} & Calibrator (K3 II-III) &2.15$\pm$0.04& 25 & 08:03-08:07 & &16.0/31.9/47.9&-107&1.1&1.3&1.3\\
45~Eri & Calibrator (K3 II-III) &2.15$\pm$0.04& 50 & 08:09-08:12 & &16.0/32.0/48.0&-107&1.1&1.3&1.3\\
\object{$\gamma$~Eri}& Check star (M0.5 IIIb)&8.74$\pm$0.09& 25 & 08:22-08:26 & &15.6/31.2/46.8&-104&1.1&1.4&1.2\\
$\gamma$~Eri& Check star (M0.5 IIIb)&8.74$\pm$0.09& 50 & 08:28-08:32 & &15.5/31.0/46.5&-104&1.1&1.4&1.2\\
S~Ori&Science target&& 25 & 08:45-08:49 & 0.78 &15.9/31.8/47.8&-107&1.1&1.3&1.3\\
S~Ori&Science target&& 50 & 08:52-08:57 & 0.78 &16.0/31.9/47.9&-107&1.1&1.3&1.3\\
$\alpha$~Hor & Calibrator (K2 III) &2.76$\pm$0.03& 25 & 09:14-09:18 &&14.5/28.9/43.3&-91&1.1&1.2&1.4\\
$\alpha$~Hor & Calibrator (K2 III) &2.76$\pm$0.03& 50 & 09:21-09-24 &&14.3/28.7/43.0&-90&1.1&1.4&1.2\\
S~Ori & Science target && 25 & 09:36-09:40 & 0.78 &15.9/31.8/47.7&-106&1.1&1.5&1.1\\\hline
\end{tabular}
\end{table*}
\section{Observations and data reduction}
We obtained near-infrared $J$, $H$, $K$ interferometry of S~Ori
with the instrument AMBER (Petrov et al. \cite{petrov07}) in low-resolution
mode at the ESO VLTI using the fringe tracker FINITO and three
VLTI Auxiliary Telescopes (ATs) on 12 October 2007 (JD 2454386).
The ATs were positioned on stations E0, G0, and H0.
The date of observation corresponds to a visual phase $\Phi_\mathrm{Vis}$=0.78,
with an uncertainty of about 0.1.
The details of the observing sequence are listed in Table~\ref{tab:obslog},
including the projected baseline lengths ($B_p$) and position angles (PA$_p$),
the airmass (AM), and the optical seeing and coherence time.
Ambient conditions were not very good but stable. The airmass
was the same for all observations. Data were
recorded using two different detector integration times (DITs) 
of 25\,msec and 50\,msec.

The VLTI fringe-tracker FINITO records fringes on the two shortest baselines
using 70\% of the $H$-band light. Output signals are processed in realtime 
and used to compensate for the fringe motion due to atmospheric turbulence. 
Owing to the low coherence time during the observations, FINITO was only 
able to provide average performances of 0.2--0.4\,$\mu$m RMS (to be compared with
0.1\,$\mu$m RMS achieved in good conditions). Nevertheless such
a performance is sufficient for increasing the signal-to-noise ratio (S/R) of 
the AMBER data and stabilizing the transfer function.
All observations used here were obtained with exactly the same
FINITO controller parameters, which is important for avoiding systematic 
biases in calibrating the absolute visibility.

In addition to S~Ori, the two calibration stars 45~Eri and $\alpha$~Hor were
observed close in time, as was $\gamma$~Eri.
The last is a regular non-pulsating M\,0.5 giant with a 
well-known angular diameter similar to that of S~Ori which is not expected to
exhibit strong effects from molecular layers. 
This data is used to check that any strong wavelength-dependent 
features found for S~Ori are not caused by any systematic effects of 
the instrument. The spectral types and angular diameters of the calibration 
and check stars are from Bord\'e et al. (\cite{borde02}). 

Raw visibility and closure phase values were computed
using the latest version of the {\tt amdlib} package (version 2.0 beta 2b)
and the {\tt yorick} interface provided by the AMBER consortium and 
the Jean-Marie Mariotti Center.
Absolute wavelength correlation was performed by correlating
the raw spectra of all four stars with a model of the atmospheric
transmission with the same spectral resolution. In particular, we used
a plateau in the transmission curve at $\lambda\sim$\,2.0\,$\mu$m. The offset
with respect to the original wavelength table was 3 spectral channels 
(0.1\,$\mu$m at a wavelength of 2.0\,$\mu$m).
We estimated the error of the absolute wavelength calibration
to 1\,pixel ($\sim$\,0.03\,$\mu$m).
Individual frames were averaged after frame selection 
keeping 70\% of the best frames based on piston (to 
remove the frames when the FINITO loop was not closed) and out of these keeping
30\% of the best frames based on S/R. We verified that keeping up to
80\% of the best frames based on SNR did not significantly change the results.
The S~Ori and $\gamma$~Eri visibility data were 
calibrated for each DIT value separately using the 45~Eri and $\alpha$~Hor 
calibration star data.
After calibration, the different calibrated S~Ori and $\gamma$~Eri data 
were averaged.
The errors of the calibrated visibility data include the statistical error 
of averaging the single frames, the errors of the calibration stars' angular 
diameters, and the variation of the available transfer function measurements.
\section{Results}
\begin{figure*}
\centering
\resizebox{0.247\hsize}{!}{\includegraphics{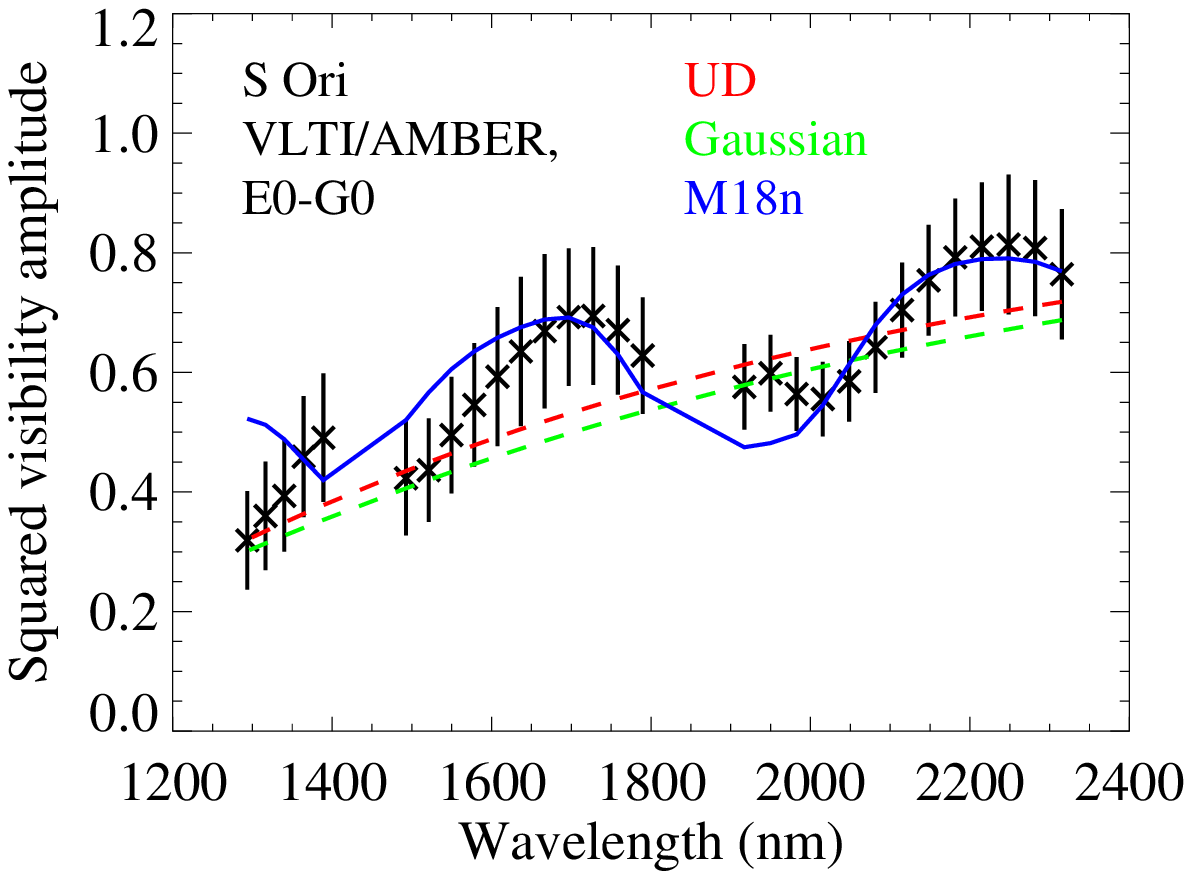}}
\resizebox{0.247\hsize}{!}{\includegraphics{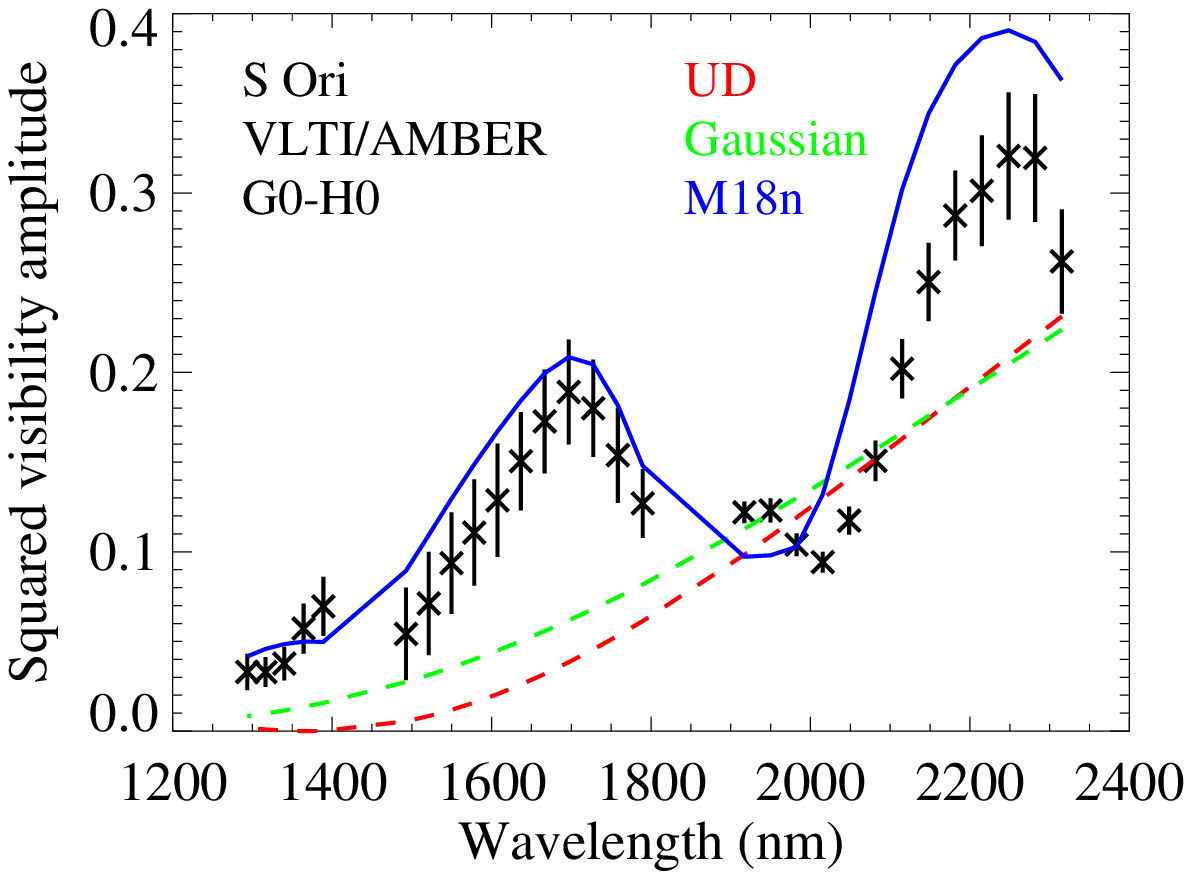}}
\resizebox{0.247\hsize}{!}{\includegraphics{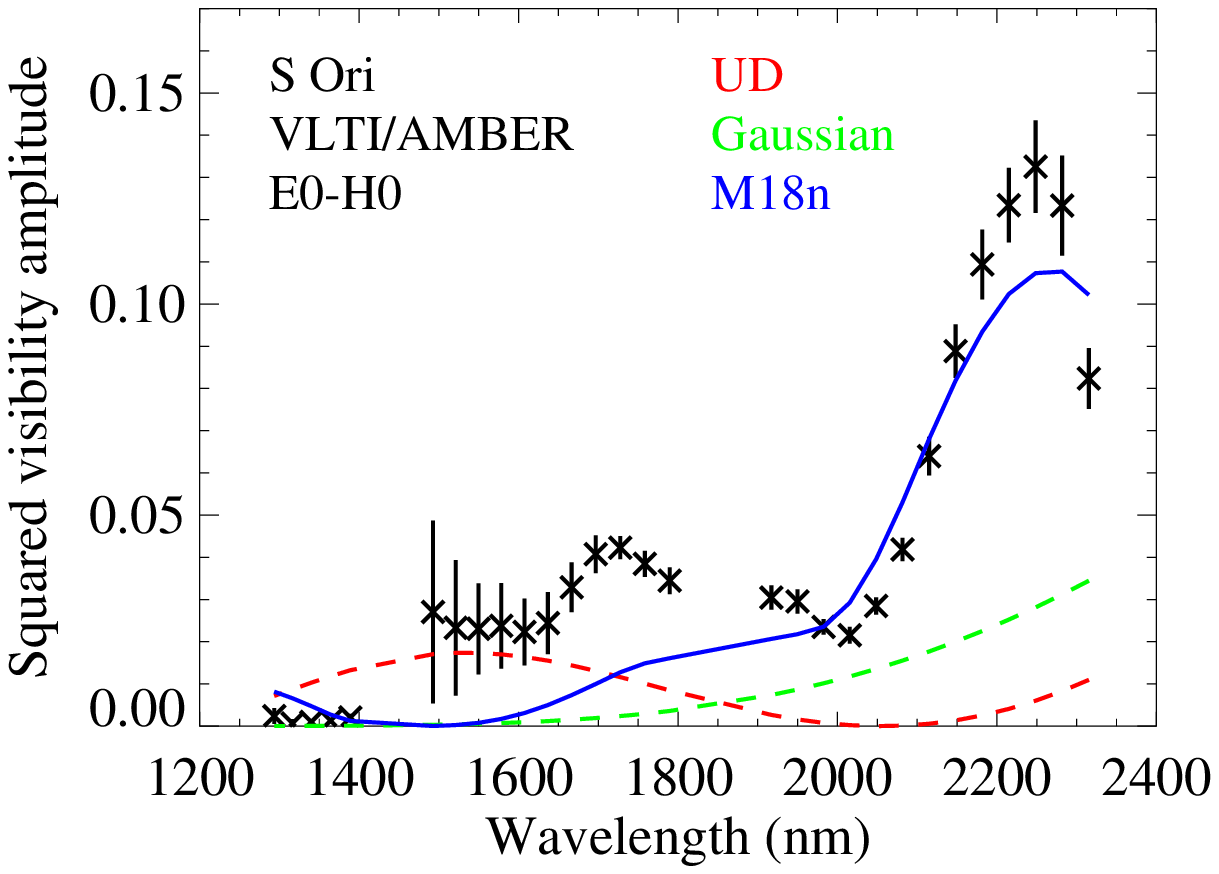}}
\resizebox{0.247\hsize}{!}{\includegraphics{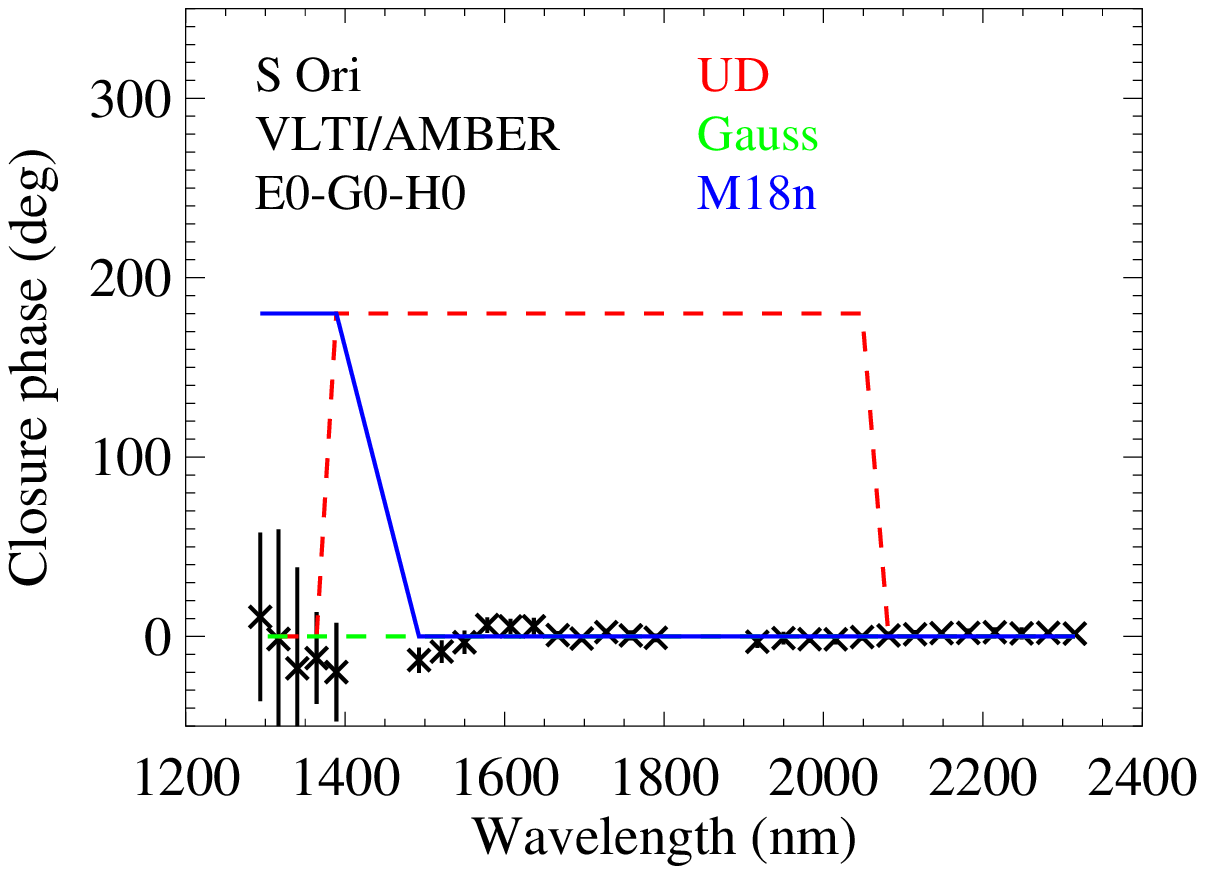}}
\caption{Measured S~Ori visibility data compared to models 
of a UD with a constant diameter (red dashed lines), 
of a Gaussian disk of constant diameter (green dashed line),
and of the M18n atmosphere model (blue solid line). For the projected
baseline lengths and angles see Table~\protect\ref{tab:obslog}.}
\label{fig:sori}
\end{figure*}
\begin{figure*}
\centering
\resizebox{0.247\hsize}{!}{\includegraphics{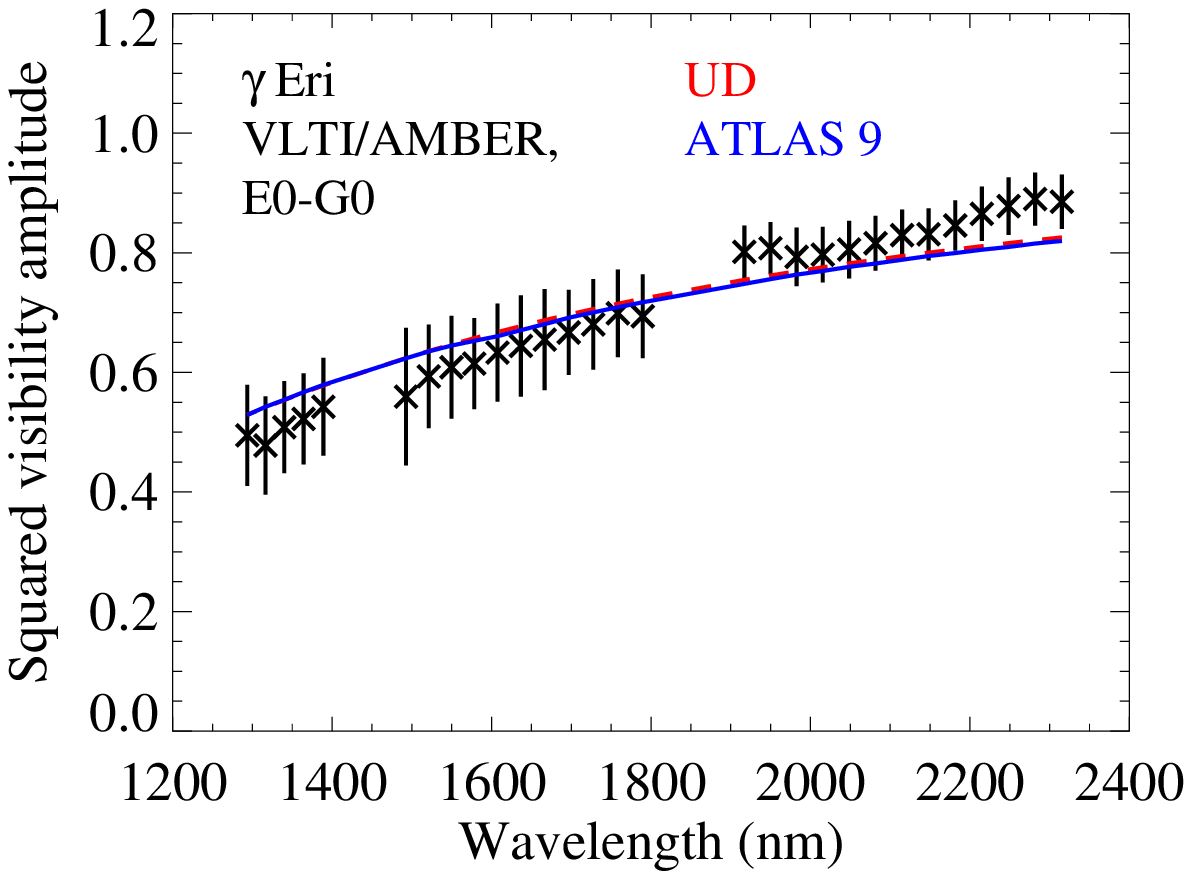}}
\resizebox{0.247\hsize}{!}{\includegraphics{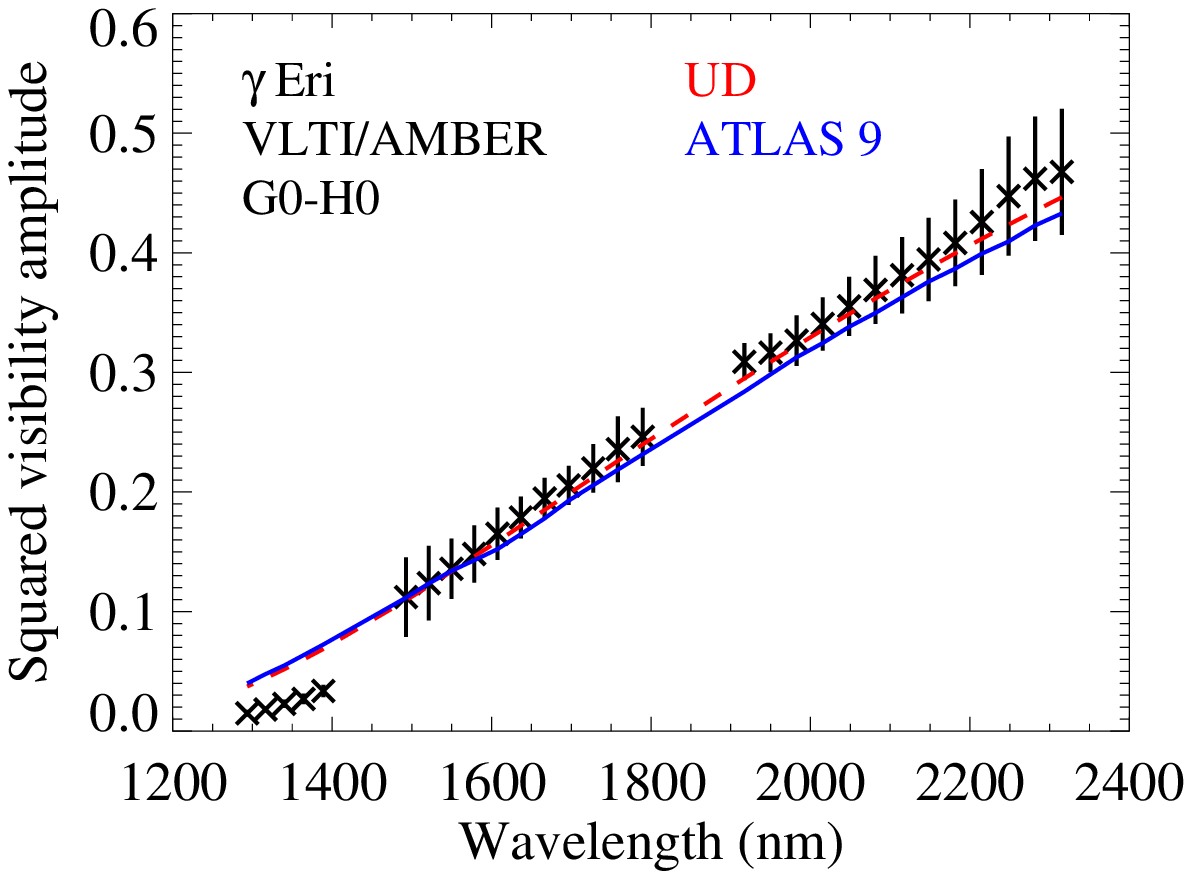}}
\resizebox{0.247\hsize}{!}{\includegraphics{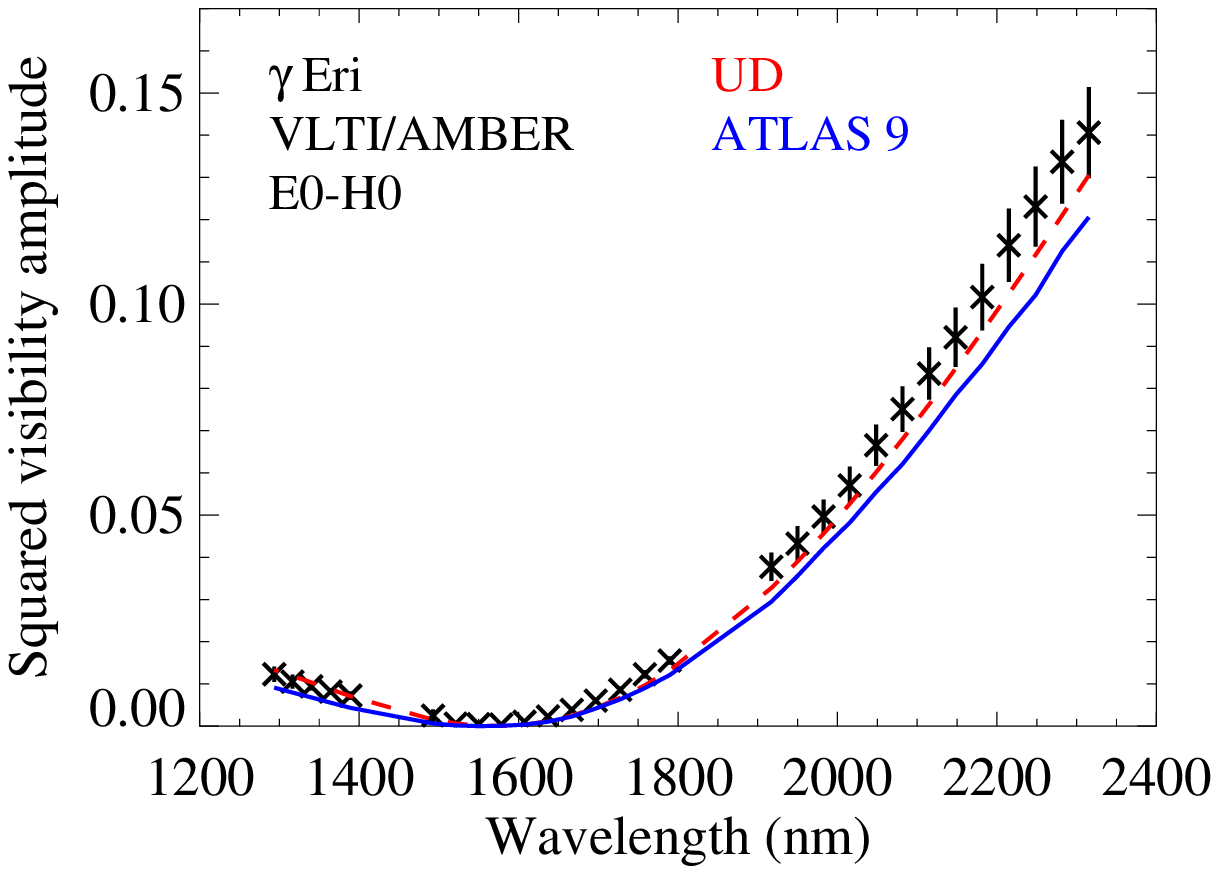}}
\resizebox{0.247\hsize}{!}{\includegraphics{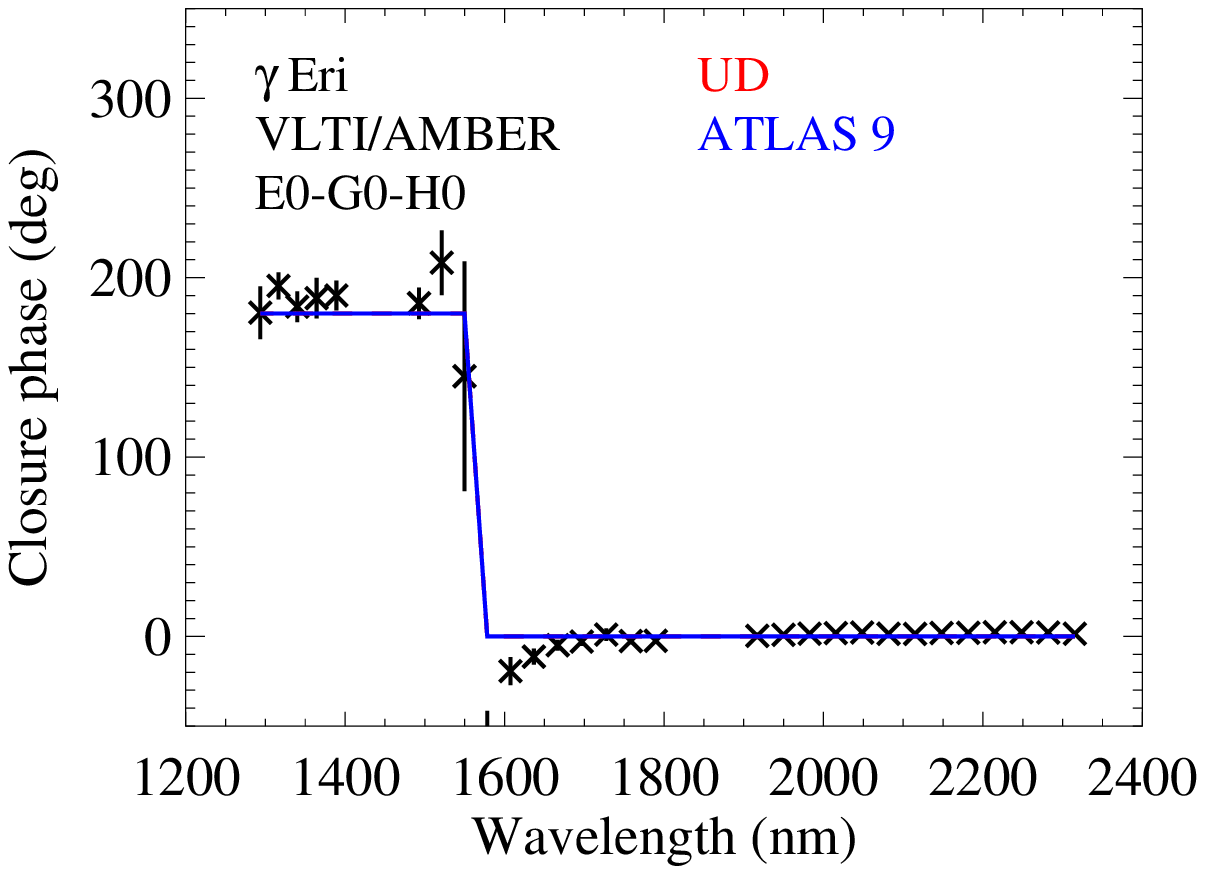}}
\caption{Measured $\gamma$~Eri visibility data compared to models
of a UD with a constant diameter (red dashed line) and of 
an {\protect\tt ATLAS\,9} model
atmosphere with $T_\mathrm{eff}$=3750\,K, $\log g$=1.5, solar chemical abundance
(blue solid line). For the projected
baseline lengths and angles see Table~\protect\ref{tab:obslog}.
}
\label{fig:gameri}
\end{figure*}
Figures~\ref{fig:sori} and \ref{fig:gameri} show the resulting visibility 
and closure phase data for S~Ori and for the check star $\gamma$~Eri, 
respectively. 
The gaps in the visibility data around 1.45\,$\mu$m and 1.85\,$\mu$m
correspond to the regions between the bands.
Also shown are the best fitting models of a UD with a constant diameter,
of a Gaussian disk with a constant diameter, and of atmosphere models.
The latter are for S~Ori the M18n model 
(described in detail below in Sect.~\ref{sec:comp}), 
and for $\gamma$~Eri an ATLAS\,9 model 
atmosphere (Kurucz \cite{kurucz}) 
with $T_\mathrm{eff}$=3750\,K, $\log g$=1.5, and solar chemical abundance 
as expected for $\gamma$~Eri (Bord\'e et al. \cite{borde02}).
The comparison of the S Ori visibility data to the dynamic model atmosphere is 
described in detail below in Sect.~\ref{sec:comp}.

The calculation of synthetic visibility values and the fits to the
interferometric data were performed as in Wittkowski et al. 
(\cite{wittkowski06,wittkowski07}). 
The fitted angular diameter of the Gaussian model corresponds to the FWHM,
that of the plane-parallel ATLAS\,9 model to the 0\% intensity (limb-darkened) 
radius, and that of the M18n model to the 1.04\,$\mu$m (photospheric) radius 
(as defined in Ireland et al. \cite{ireland04c}). 
The best-fit angular diameters are\\
S~Ori: $\Theta_\mathrm{UD}$=10.8\,mas; $\Theta_\mathrm{Gaussian}$=6.9\,mas;
$\Theta_\mathrm{M18n}$=8.3\,mas\\
$\gamma$~Eri: $\Theta_\mathrm{UD}$=8.5\,mas; $\Theta_\mathrm{ATLAS 9}$=8.9\,mas.\\
The errors are $\sigma\sim$\,0.2\,mas

The visibility data of the check star $\gamma$~Eri can be described well
by a UD of constant diameter and by the {\tt ATLAS\,9} model
atmosphere. There are no significant wavelength-dependent deviations between
measured visibility data and the UD model. It is not yet
clear whether the relatively low $J$-band 
visibilities for the G0-H0 baseline and the deviations in the 
closure phase values near the flip are caused by an asymmetric stellar 
surface feature 
or a systematic calibration uncertainty.
Note that the angular diameter of $\gamma$~Eri based on a given model is 
well-constrained by the position of the visibility minimum, 
which is independent of an absolute visibility calibration. The resulting
angular diameter $\Theta_\mathrm{ATLAS 9}$=8.9$\pm$0.2\,mas
is consistent with the value given in Bord\'e et al. (\cite{borde02}) 
of $\Theta_\mathrm{LD}$=8.74$\pm$0.09\,mas.

The visibility data of S~Ori show significant wavelength-dependent 
features clearly deviating from UD and Gaussian models of constant diameter
on all three baselines. This indicates variations in the apparent
angular diameter. The S~Ori closure phase values are consistent with zero at
all spectral channels, indicating the absence of asymmetric features
in the intensity distribution.
\begin{figure}
\centering
\resizebox{0.86\hsize}{!}{\includegraphics{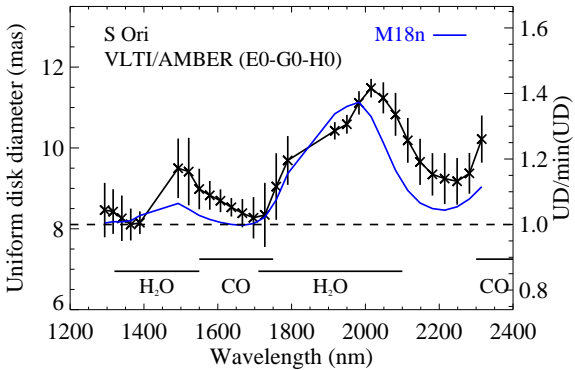}}
\caption{S~Ori UD diameter values as a function of wavelength
compared to the prediction by the M18n model atmosphere. 
Also indicated are the positions of H$_2$O and CO bands after
Lan{\c c}on \& Wood (\cite{lancon00}).}
\label{fig:soriud}
\end{figure}
In order to characterize the variation of S~Ori's angular diameter
as a function of wavelength, we fitted UD diameters
to the data of each spectral channel separately. 
Figure~\ref{fig:soriud} shows the
resulting UD diameter values as a function of wavelength. 
Note that the intensity 
profile of a Mira star is generally not expected to be close to a UD and 
that this approach can only give a rough estimate of S~Ori's characteristic size
as a function of wavelength. Fits of Gaussian functions lead to similar results.
The apparent UD angular diameter shows clear variations with wavelength.
It is smallest at about 1.3\,$\mu$m and 1.7\,$\mu$m and increases by
a factor of $\sim$\,1.4 around 2.0\,$\mu$m.
The minimum UD angular diameter of S~Ori at the near-continuum 
wavelengths is $\Theta_\mathrm{UD}$=8.1$\pm$0.5\,mas. 
This result at visual phase 0.78 is consistent with the S~Ori 
photospheric angular diameters 
between 7.9\,mas at phase 0.55 and 9.7\,mas at phase 1.15 derived
in Wittkowski et al. (\cite{wittkowski07}) based on VLTI/MIDI data
and modeling with dynamic model atmospheres and a radiative transfer
model of the dust shell. These photospheric angular diameters are
also consistent with previous broadband measurements 
(cf. Wittkowski et al. \cite{wittkowski07}). 
With the adopted distance to S~Ori, the angular diameter 
$\Theta_\mathrm{UD}$=8.1$\pm$0.5\,mas corresponds to a radius
418\,$\pm$\,130\,R$_\odot$.
\section{Comparison to dynamic Mira star atmosphere models}
\label{sec:comp}
\begin{figure*}
\centering
\resizebox{0.33\hsize}{!}{\includegraphics{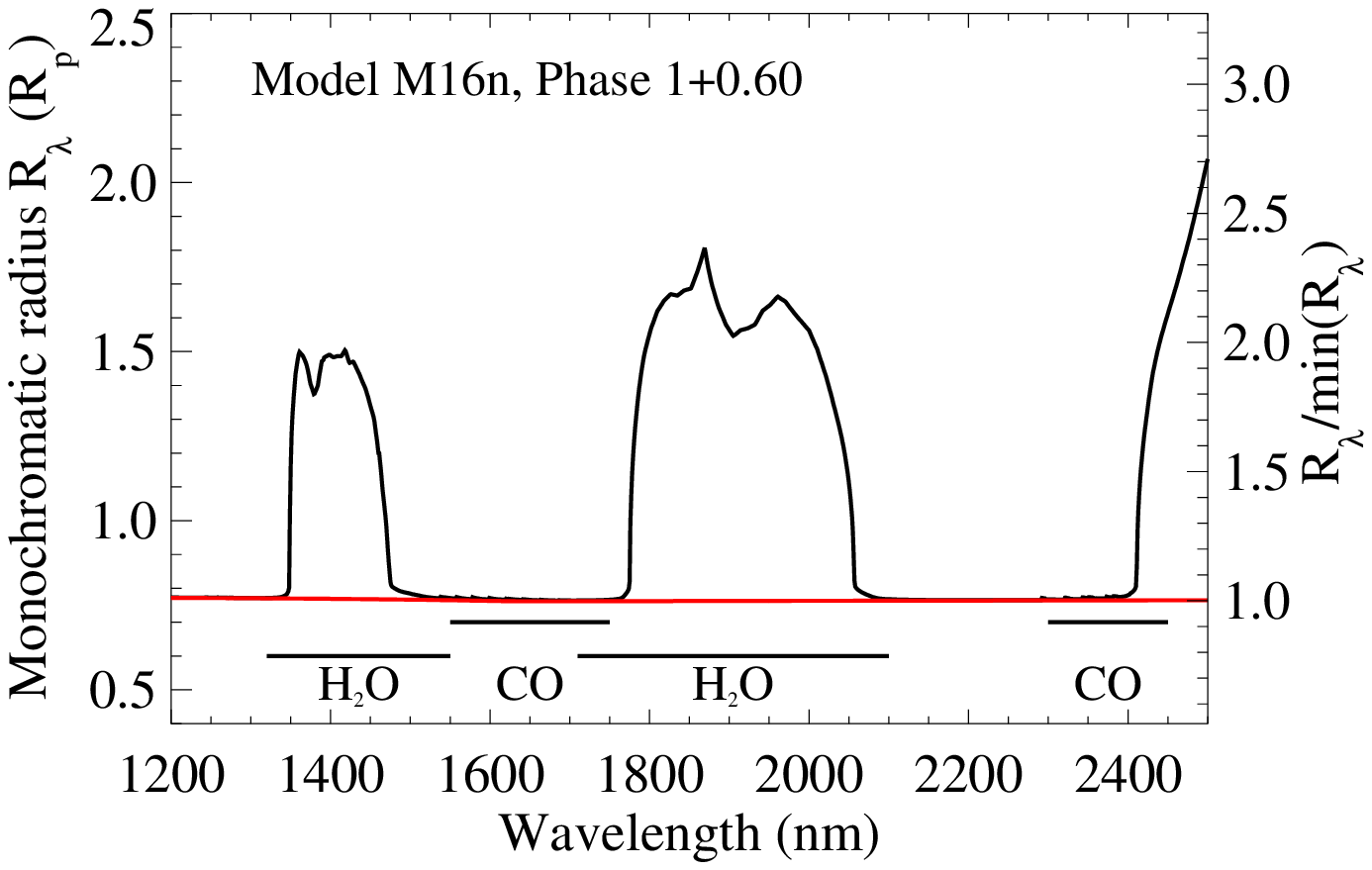}}
\resizebox{0.33\hsize}{!}{\includegraphics{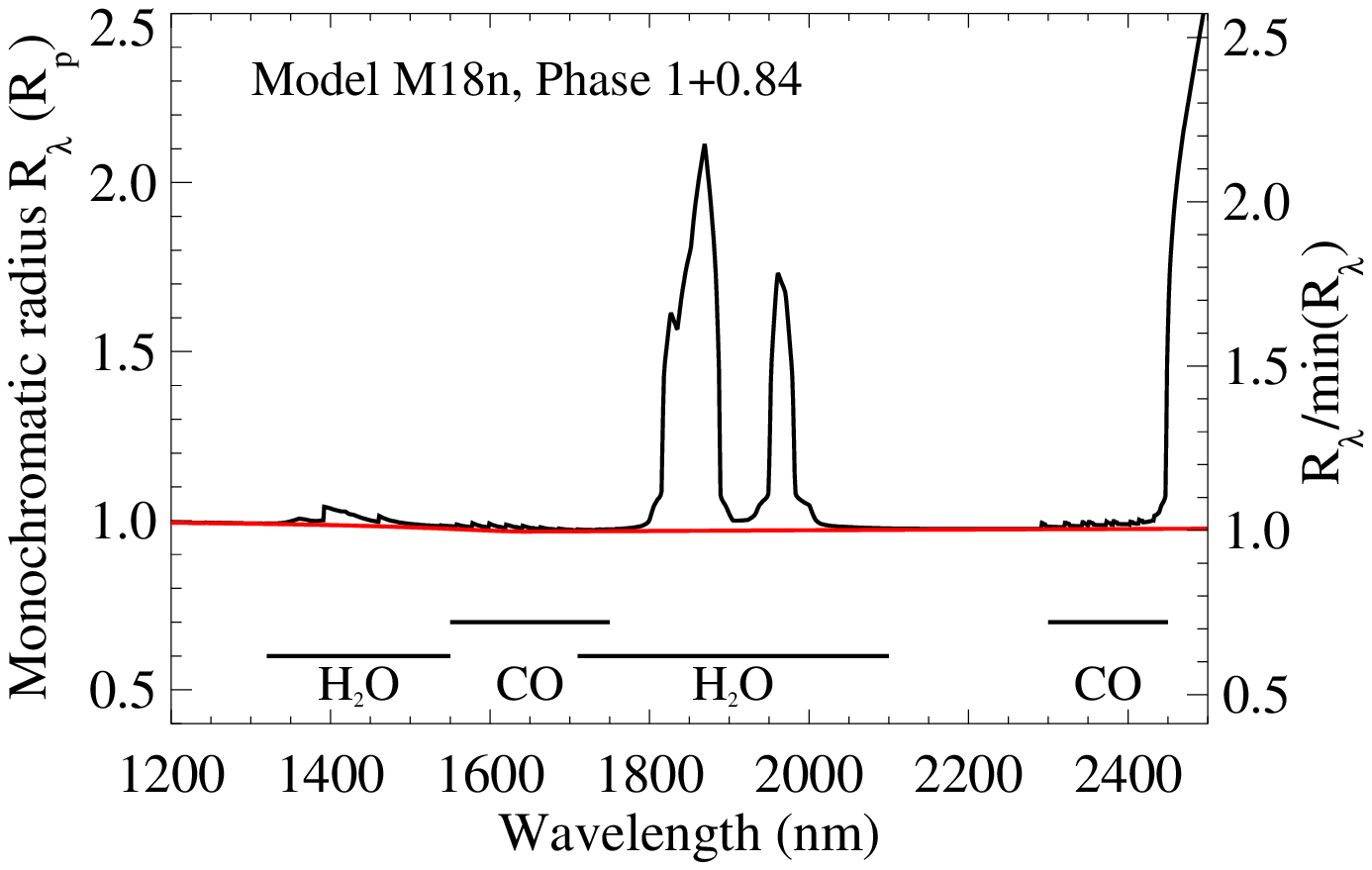}}
\resizebox{0.33\hsize}{!}{\includegraphics{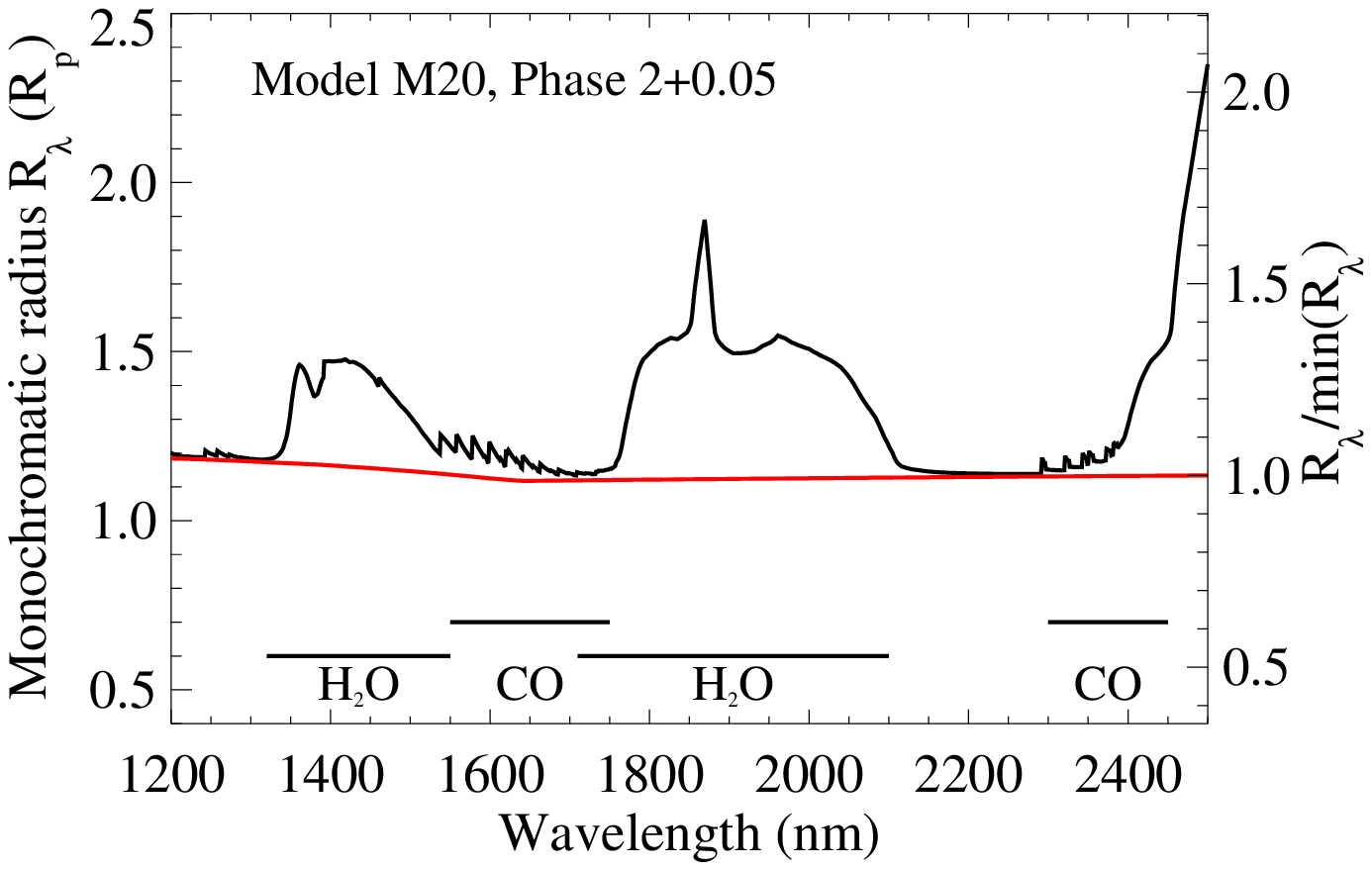}}
\caption{Monochromatic $\tau_\lambda=1$ Mira star radii predicted 
by the M model series for the example of the M16n (model phase 1.60), 
M18n (1.84), and M20 (2.05) models. The red line indicates the model continuum
radius excluding atomic and molecular features. The spectral resolution 
of the model is 0.001\,$\mu$m.
Also indicated are the positions of H$_2$O and CO bands after
Lan{\c c}on \& Wood (\cite{lancon00}).}
\label{fig:Mmodels}
\end{figure*}
Few dynamic atmosphere models for oxygen-rich Mira stars
are available that include the effects of molecular
layers. The P and M model series (Ireland et al. \cite{ireland04b,ireland04c})
are complete self-excited dynamic 
model atmospheres of Mira stars designed to match the prototype
oxygen-rich Mira stars $o$~Cet and R~Leo. They have been used successfully
for comparisons to recent broadband interferometric data of $o$~Cet and R~Leo 
(Woodruff et al. \cite{woodruff04}; Fedele et al. \cite{fedele05}).
Compared to $o$~Cet and R~Leo, S~Ori is a slightly
cooler Mira variable with a longer period, a higher main-sequence precursor
mass, and a larger radius. However, when scaled to variability phases 
between 0 and 1 and to the corresponding angular size on the sky, the 
general model results are not expected to be dramatically different for
S~Ori compared to $o$~Cet and R~Leo (cf. the discussion
in Wittkowski et al. \cite{wittkowski07}). The M model series
was chosen to model the atmosphere of S~Ori by 
Wittkowski et al. (\cite{wittkowski07}) as the currently best available 
option to describe Mira star atmospheres.
Monochromatic center-to-limb variations (CLVs) at 46 angles between 0 and 5 $R_p$
based on the P and M models were recomputed
for the wavelength range from 1--2.5\,$\mu$m 
in steps of 0.001\,$\mu$m.

The P and M dynamic model atmospheres predict significant changes in the 
monochromatic radius $R_\lambda=R_\lambda(\tau_\lambda=1)$,
caused by molecular layers that lie above the 
continuum-forming photosphere and significantly affect certain bandpasses.
Figure~\ref{fig:Mmodels} shows 
the monochromatic radius $R_\lambda=R_\lambda(\tau_\lambda=1)$ in units
of the non-pulsating parent star radius for the example of three
models of the series. 
It illustrates the strong phase
dependence of the molecular layers. 
The models are
M16n (model phase 1.60), M18n (1.84), and M20 (2.05) models. 
The red lines denote
for comparison the $R_\lambda$ values solely based on the continuum 
radiation excluding all atomic and molecular features.
Also shown are the positions of the H$_2$O and CO bands after
Lan{\c c}on \& Wood (\cite{lancon00}) and references therein.
The most prominent features of these model curves in the near-infrared 
region
are two water vapor features
around 1.4\,$\mu$m and 1.9\,$\mu$m, and also CO features
around 1.6\,$\mu$m and 2.4\,$\mu$m.
The strengths, shapes, and widths of these molecular
features depend strongly on the stellar phase (and also on cycle), as is
evident from the comparison of the three model curves. Also, the relative
strengths of the molecular features varies with stellar phase. 

Model M18n provides the best formal fit to our S~Ori AMBER visibility
data out of the 20 available phase and cycle combinations of the M series. 
The synthetic visibility values based on the M18n model 
compared to our AMBER observation
are indicated in Fig.~\ref{fig:sori}. Here, the angular photospheric diameter
corresponding to the 1.04\,$\mu$m (photospheric) model radius 
(defined in Ireland et al. \cite{ireland04c}) is 
$\Theta_\mathrm{M18n}$=8.3\,$\pm$\,0.2\,mas, consistent
with the UD diameter at near-continuum wavelengths 1.3\,$\mu$m and 1.7\,$\mu$m
of $\Theta_\mathrm{UD}$=8.1$\pm$0.5\,mas.
The theoretical $R_\lambda(\tau_\lambda=1)$ radii in Fig.~\ref{fig:Mmodels}
cannot be compared directly to the UD diameters derived from our 
AMBER data (Fig.~\ref{fig:soriud}), because of the 
different spectral resolution and because the model-predicted
CLVs can be very different from a UD model (so that different
radius definitions are not equal).
The translation of the model prediction into a UD diameter depends on the 
exact shape of the bandpass-averaged CLV and the baselines used.
To compare the model predictions to the measured UD values
in Fig.~\ref{fig:soriud}, we
fitted UD diameters to the synthetic visibility values of the M18n model 
using exactly the same spectral channels, baseline configuration, and 
fit method as for our AMBER data. The resulting model prediction for the 
UD diameter as a function of wavelength is shown in Fig.~\ref{fig:soriud},
in comparison to the values derived from the observation.

Figures~\ref{fig:sori} and \ref{fig:soriud} show that our AMBER visibility
data can be described reasonably well by the dynamic atmosphere model
M18n. The differences between observations and model prediction are 
most likely due to an imperfect match of the phase and cycle combination
between observation and available M models of the series. Looking
at Figs.~\ref{fig:soriud} and \ref{fig:Mmodels}, it is evident 
that a better fit to our AMBER data could be obtained with a model that
shows a stronger 1.4\,$\mu$m water vapor feature of the same shape
compared to M18n (as seen for instance in the case of M20), 
and at the same time a just as strong but broader 1.9\,$\mu$m feature (as for 
instance in the case of M16n).
It is quite possible that such a combination of the two water-vapor features 
could appear for a model of another phase-cycle combination.
Also, some differences between M model predictions and observations of
S~Ori are expected due to the different stellar parameters of S~Ori
compared to the parent star of the M model series. Finally, differences
can also be caused by remaining uncertainties in the absolute calibration
of visibility values and of the wavelength scale. The S~Ori dust shell 
of $\tau_V$=1.5--2.5 as modeled in Wittkowski et al. (\cite{wittkowski07}) 
is not expected to have a noticeable effect on our near-infrared visibility data, 
because its contribution to the visibility was already small at 8\,$\mu$m 
and because the visibility data are calibrated separately for each 
spectral channel.\\

In summary, our AMBER observations of S Ori generally confirm
the predictions by the M model series and we find that the observed variation of 
diameter with
wavelength can be understood as the effect of phase-dependent water vapor
and CO layers lying above the photosphere. The M model series can
be used reasonably well to model the atmosphere of a Mira star such as S Ori
and to derive a reliable photospheric radius based on broadband data. 
More such observations are needed to 
confirm and constrain the model predictions in more detail and to monitor 
the predicted phase dependence of the strength and characteristics of the 
molecular layers. Simultaneously obtained spectra would be a valuable addition.
\begin{acknowledgements}
We acknowledge with thanks the use of the AMBER data reduction
software from the JMMC (version 2.0 beta 2b). 
\end{acknowledgements}
\end{document}